\documentclass[prb,preprint,aps]{revtex4}
\usepackage{graphicx}
\usepackage{epstopdf}
\newcommand{\dt}{{\Delta t}}
\newcommand{\e}{{\rm e}}

\newcommand{\p}{{\bf p}}
\newcommand{\bv}{{\bf v}}
\newcommand{\ac}{{\bf a}}

\newcommand{\br}{{\bf r}}
\newcommand{\q}{{\bf q}}
\newcommand{\f}{{\bf F}}

\newcommand{\bea}{\begin{eqnarray}}
\newcommand{\eea}{\end{eqnarray}}
\newcommand{\be}{\begin{equation}}
\newcommand{\ee}{\end{equation}}
\newcommand{\ba}{\begin{eqnarray}}
\newcommand{\ea}{\end{eqnarray}}

\newcommand{\nn}{\nonumber}
\newcommand{\la}{\label}

\def\t1{e_{_T}}
\def\v1{e_{_V}}

\begin{document}
%\tightenlines
\title{Newton's graphical method as a canonical transformation }

\author{Siu A. Chin}

\affiliation{Department of Physics and Astronomy, 
Texas A\&M University, College Station, TX 77843, USA}

%\date{\today}
\begin{abstract}
This work shows that, Newton's Proposition 1 in the {\it Principia},
is an {\it exact} graphical representation of a canonical transformation,
a first-order symplectic integrator generated at a finite time-step by 
the Hamiltonian. A fundamental characteristic of this canonical transformation is to
update the position and velocity vectors {\it sequentially}, thereby
automatically conserving the phase-volume and the areal velocity
due to a central force. As a consequence, the continuous force is naturally 
replaced by a series of impulses.  
The convergence of Newton's Proposition 1 in the limit of
$\dt\rightarrow 0$ can be proved easily and the resulting error term
for the linear and the inverse square force can 
explain why Hooke was able to the obtain an elliptical 
orbit for the former but not the latter.
\end{abstract}
\maketitle

\section {Introduction}
\la{intro}
									 
Newton's Proposition 1, Theorem 1, in Book I of his {\it Principia}, has always
been admired for its effortless demonstration of Kepler's area law for {\it any} 
central force. Recently in this Journal, Nauenberg\cite{nau18} transcribed Newton's
graphical construction into an algebraic algorithm, cited existing algorithms\cite{storm,ver67,cro81}
as unaware of their connections to Newton's Proposition 1 and credited
Ref.\cite{cou04} for explicitly identifying Newton's Proposition 1 with a particular algorithm.
This is very surprising, since Nauenberg himself has correctly identified the algorithm corresponding
to Proposition 1 ten years earlier!\cite{nau94}

It seems clear that even after identifying Newton's Proposition 1 with an algebraic algorithm, 
the authors of Ref.\onlinecite{cou04} and Nauenberg\cite{nau94}, were unaware of the full extent of
this identification implies. While Nauenberg\cite{nau94} mentioned ``canonical or symplectic transformation''
in his Ref.74 there, that reference did not contain the algorithm as an example. (Also, the transformation is
not time dependent.) This work first show that the algorithm is one of two fundamental canonical 
transformations generated by the Hamiltonian. As such, it automatically conserve phase-volume\cite{gol80,lan60} with unit determinant and maintains the constancy of the areal velocity due to any central force.
Secondly, this work shows that the algorithm can also be derived as a symplectic integrator, 
with known error Hamiltonians. These error Hamiltonians can precisely quantify 
the algorithm's behavior under different central forces.
This then allows us to understand one important incident in the history of mechanics.

In an earlier work\cite{nau94}, Nauenberg has woven together a compelling account of
how Hooke has written to Newton, and told Newton of his idea
 ``of compounding the celestiall motions of the planetts of a direct motion by the tangent and an attractive motion toward the central body...'' in a letter\cite{nau94} dated November 14, 1679.  
Nauenberg suggested that Newton may have absorbed Hooke's idea in formulating his Proposition 1. Of course, Newton's Proposition 1 is mathematically more precise than Hooke's mere description, and its proof of Kepler's second law owed nothing to Hooke. But the story continued with Hooke: upon learning of Newton's graphical construction, he immediately applied it to the case of a linearly increasing central force and obtained an elliptical orbit! However, there is no indication that Hooke was ever successful in obtaining an elliptical orbit for the inverse square force.\cite{nau94} 
This work will explain why in Section \ref{error}. 

In the next section we identify the two fundamental canonical transformations 1A and 1B generated by the Hamiltonian of the system. In Section \ref{newton} we show that Newton's Proposition 1 is an exact graphical representation of transformation 1B.  In Section \ref{sym}, we dervie 1B as a symplectic algorithm and 
resolved questions about the convergence and time-reversibility of Newton's proposition 1.
In Section \ref{error}, we show that the error Hamiltonian associated with 1B is very different for a linear and an inverse square force, thereby explaining why Hooke can obtain an elliptical orbit for the former, but not the latter. A brief concluding statement is given in Section \ref{con}.

\section {Canonical transformation}
\la{cano}

A canonical transformation $(q_i,p_i)\rightarrow (Q_i,P_i)$ is a transformation which
preserve the form of Hamilton's equations for both sets of variables.\cite{gol80, lan60}
Canonical transformations can be derived on the basis of 
four types of {\it generating} functions,\cite{gol80, lan60}
$F_1(q_k,Q_k,t)$, $F_2(q_k,P_k,t)$, $F_3(p_k,Q_k,t)$, $F_4(p_k,P_k,t)$. 
For our purpose, we will
only need to use $F_2$ and $F_3$ without any explicit time-dependence,
given by
\be
p_i=\frac{\partial F_2(q_k,P_k)}{\partial q_i},\qquad 
Q_i=\frac{\partial F_2(q_k,P_k)}{\partial P_i},
\la{f2}
\ee
and
\be
q_i=-\frac{\partial F_3(p_k,Q_k)}{\partial p_i},\qquad 
P_i=-\frac{\partial F_3(p_k,Q_k)}{\partial Q_i}.
\la{f3}
\ee
In (\ref{f2}),
the first equation is an implicit
equation for finding $P_i$ in terms of $q_i$ and $p_i$. 
The second, is an explicit equation for determining $Q_i$ in 
terms of $q_i$ {\it and} the updated $P_i$. This can be viewed
naturally as a {\it sequence} of two transformations: first to $P_i$ then $Q_i$.
This sequential updating of  $P_i$ and $Q_i$ 
automatically guarantees the preservation of
phase-volume,\cite{gol80,lan60} so that the determinant of the Jacobian of 
transformation is one. Similarly for $F_3$, but now first transform to $Q_i$ then $P_i$.
This sequential updating of $Q_i$ and $P_i$ is a hallmark of canonical  transformations 
and naturally leads to the replacement of a continuous force by a series of impulses. 

Among canonical transformations, the most important one is when
$Q_i\!=\!q_i(t)$ and $P_i\!=\!p_i(t)$, which solves the dynamics of
the system. For an arbitrary $t$, the required transformation is generally 
unknown. However, when $t$ is infinitesimally
small, $t\rightarrow \dt$, it is well known that the Hamiltonian is the {\it infinitesimal }  
generator of time evolution.\cite{gol80,lan60}  Less well known is the fact that
even when $\dt$ is {\it finite}, the resulting transformation generated by the Hamiltonian
remains canonical, and gives an excellent {\it approximate} trajectory. Let's take 
\be
F_2(q_i,P_i)=\sum_{i=1}^n q_iP_i+\dt H(q_i,P_i)
\la{f22}
\ee
where 
\be
H(q_i,p_i)=\sum_{i=1}^n\frac{p_i^2}{2m}+V(q_i)
\la{ham}
\ee
is the usual separable Hamiltonian.
For this generating function (\ref{f22}), $\dt$ is simply an arbitrary parameter, need not be small. 
The transformation equations (\ref{f2}) then give,
\be
P_i=p_i-\dt\frac{\partial V(q_i)}{\partial q_i},\qquad Q_i=q_i+\dt\frac{P_i}m
\la{alf2} 
\ee
If one regards $q_i=q_i(t)$, $p_i=p_i(t)$ and $Q_i=q_i(t+\dt)$, $P_i=p_i(t+\dt)$ 
then the above is precisely the algorithm that Cromer's student rediscovered accidentally\cite{cro81}. 
For uniformity, we can simply referring to this as algorithm 1A.
Similarly, taking
\be
F_3(p_i,Q_i)=-\sum_{i=1}^n p_iQ_i+\dt H(Q_i,p_i)
\ee
and applying (\ref{f3}) gives the other canonical algorithm
\be
Q_i=q_i+\dt\frac{p_i}m,\qquad P_i=p_i-\dt\frac{\partial V(Q_i)}{\partial Q_i}.
\la{alf3}  
\ee
This algorithm has no name, but was listed by Stanley\cite{sta84} among others.\cite{cou04,nau94}
We will refer to this algorithm as 1B. 
Among the many elementary algorithms studied, such as those by Stanley\cite{sta84}, 
only these two, and their variants, were later found to be explicit, self-starting
and symplectic.
For the ease of comparison, we will define $\br_n=\q(n\dt)$, $\bv_n\equiv \p(n\dt)/m$,  
$\ac(\br)\equiv \f(\br) /m=-\nabla V(\br)/m$. The above two algorithms can then be succintly 
given in modern notations as 1A:
\ba
\bv_{n+1}&=&\bv_n+\dt \ac(\br_n)\nn\\
\br_{n+1}&=&\br_n+\dt \bv_{n+1}
\la{cor}
\ea
and 1B:
\ba
\br_{n+1}&=&\br_n+\dt \bv_{n}\la{new}\\
\bv_{n+1}&=&\bv_n+\dt \ac(\br_{n+1}).
\la{new2}
\ea
Note the sequential updating nature of these two algorithms. In 1A, the updated
$\bv_{n+1}$ is immediately used to compute $\br_{n+1}$. In 1B, the updated
$\br_{n+1}$ is immediately used to compute $\bv_{n+1}$.  
Because of this, not only is phase volume preserved, angular momentum is also automatically 
conserved for a central force $\ac(\br)=f(r)\br$ 
at each time step. For example, for algorithm 1B
\ba
\br_{n+1}\times\bv_{n+1}&=&\br_{n+1}\times(\bv_n+\dt \ac(\br_{n+1}))\nn\\
&=&\br_{n+1}\times \bv_n =\br_n\times \bv_n.
\la{ang}
\ea
Similarly for algorithm 1A, just substitute in the {\it last} updated variable first in
the $\br\times\bv$ computation. In 1994, Nauenberg\cite{nau94} has already identified Newton's 
proposition 1, as used by Hooke, as algorithm 1B. (See Eq.(1) and (2) in Ref.\cite{nau94}.) He also showed that it
conserves angular momentum as in (\ref{ang}).

Other well known algorithms are simply variants of these two. For example,
if one changes the label $n$ in $\bv_n$ to $\bv_{n-\frac12}$ in 1A, and to $\bv_{n+\frac12}$ in 1B, then one has the leap-frog algorithm. Also, by doing a ``one and half" 1B from $\br_{n-1}$ one gets
\ba
\br_{n}&=&\br_{n-1}+\dt \bv_{n-1}\nn\\
\bv_{n}&=&\bv_{n-1}+\dt \ac(\br_{n})\nn\\
\br_{n+1}&=&\br_n+\dt \bv_{n}\la{ohalf}\\              
     &=&\br_n+\dt \bv_{n-1}+\dt^2\ac(\br_{n})\nn\\
     &=&\br_n+ \br_{n}-\br_{n-1} +\dt^2\ac(\br_{n}).\la{ver}
\ea
While (\ref{ohalf}) is the result of the ``one and half" 1B algorithm, eliminating
the intermediate velocity $\bv_{n-1}$ gives (\ref{ver}),
which is the Verlet\cite{ver67} algorithm. This velocity elimination fundamentally
changes algorithm. The ``one and half" 1B algorithm is completely time-reversible
for every position. Starting with (\ref{ohalf}), change $\dt\rightarrow -\dt$ 
and iterate the algorithm again, one backtracks step-by-step back to $\br_{n-1}$:
\ba
\widetilde\br_{n+1}&=&\br_{n+1}-\dt \bv_{n}=\br_n\nn\\
\widetilde\bv_{n+1}&=&\bv_{n}-\dt \ac(\br_{n})=\bv_{n-1}\nn\\
\widetilde\br_{n+2}&=&\br_n-\dt \bv_{n-1}=\br_{n-1}.\nn              
\ea
In the Verlet form (\ref{ver}), the algorithm is time-reversible
only between $\br_{n+1}$ and $\br_{n-1}$, but not between every 
successve position! By eliminating $\bv_{n-1}$, the algorithm is no longer
self-starting, and requires two starting positions $\br_{0}$ and $\br_{1}$.
All even positions would time reverse back to $\br_{0}$ and all odd positions
time reverse back to $\br_{1}$, as if two trajectories are running in parallel.
If Verlet's $\br_{1}$ coincide with 1B's $\br_{1}$,
then both will yield the same trajectory. If Verlet's $\br_{1}$
is very different from that of 1B, then very different trajectories can result.
Thus Verlet is {\it not} the same as 1B. 

By identical manipulations, one can also derive Verlet from ``half and one'' 1A algorithm,
again going $\br_{n-1}$ to $\br_{n+1}$.
%\ba
%\br_{n}&=&\br_{n-1}+\dt \bv_{n}\nn\\
%\bv_{n+1}&=&\bv_n+\dt \ac(\br_n)\nn\\
%\br_{n+1}&=&\br_n+\dt \bv_{n+1}\nn\\
%\br_{n+1}&=&\br_n+\dt \bv_{n}+\dt^2\ac(\br_{n})\nn\\
%     &=&\br_n+ \br_{n}-\br_{n-1} +\dt^2\ac(\br_{n}).\nn
%\ea
Thus both 1A and 1B can give rises to Verlet, and from Verlet, one can infer
either 1A or 1B as the underlying algorithm.

\section {Newton's graphic construction}
\la{newton}

When one examine the diagram of Newton's Proposition 1 (see Fig.\ref{prop1}), one is immediately
struck by its ingenius construction. The area sweep out by each point
of the orbit is guaranteed to be equal, when the impulse for determine that point is computed at 
the radial direction of the {\it preceding} point. Also, starting at an initial position $A$,
every successive position $B$, $C$, $D$, $E$, $F$ on the orbit receives an 
impulse, except $B$. Let $A$, $B$, $C$, $D$, etc., 
be denoted by $\br_0$, $\br_1$, $\br_2$, $\br_3$ etc., and write out
the first few iterations of algorithm 1B (\ref{new2}).
\ba
&&\br_{1}=\br_0+\dt \bv_{0},\qquad \bv_{1}=\bv_0+\dt \ac(\br_{1})\la{vone}\\
&&\br_{2}=\br_1+\dt \bv_{1},\qquad \bv_{2}=\bv_1+\dt \ac(\br_{2})\nn\\
&&\quad\ =\br_0+2\dt \bv_{0}+\dt^2 \ac(\br_{1})\la{p2}\\
&&\br_{3}=\br_2+\dt \bv_{2},\qquad\bv_{3}=\bv_2+\dt \ac(\br_{3})\nn\\
&&\quad\ =\br_2+\dt \bv_{1}+\dt^2 \ac(\br_{2})\la{p3}
%&&\br_{4}=\br_3+\dt \bv_{3},\qquad\bv_{4}=\bv_3+\dt \ac(\br_{4})\nn\\
%&&\quad\ =\br_3+\dt \bv_{2}+\dt^2 \ac(\br_{3})\la{p4}
\ea
We can refer to position vectors above and Newton's diagram simultaneously
by the dual notation $\br_0(A)$, $\br_1(B)$, etc..
Comparing the diagram with the above positions, one sees that
starting at the initial position $\br_0(A)$, $\br_1(B)$ is just a distance $\dt \bv_{0}$ 
away with no impulse. This is just the Newton's First Law.
At $\br_1(B)$, its continuing ``tangential" velocity $\bv_{0}$ (in the direction of BC) 
is ``compounded" by the central impulse at the radial direction (BV) of $\br_1(B)$, 
giving rise to $\bv_1$ as stated in (\ref{vone}). This is then Newton's Second Law. 
Now one repeats Newton's First
Law and arrived at $\br_2(C)$ from $\br_{1}(B)$. This is the path ABC in Newton's diagram.
By expanding out $\br_2(C)$ in (\ref{p2}), one sees
that it can also be viewed as arriving from $\br_0(A)$ after $2\dt \bv_{0}$
plus an impulse displacement of $\dt^2 \ac(\br_{1})$ computed at the {\it radial direction}
of the preceding position $\br_{1}(B)$. This is the path AcC in Newton's diagram.
One then repeat the Second Law to obtain $\bv_2$, then the First Law to get  $\br_3(D)$,
etc..Thus positions $A$, $B$, $C$, $D$, etc., on Newton's diagram, exactly
match positions  $\br_0$, $\br_1$, $\br_2$, $\br_3$, etc., generated
by algorithm 1B. In Fig.1 the first few numerical positions of 1B with a constant central force
is compared to Newton's diagram for Proposition 1 .
The numerical positions clearly resemble those in Newton's carefully constructed diagram.
Newton's Proposition 1 is therefore an {\it exact} pictorial representation of algorithm 1B
and 1B should be named after Newton because it is simply the repeated applications of his
First and Second laws.

\begin{figure}
	%	\vspace{0.5truein}
	\includegraphics[width=0.40\linewidth]{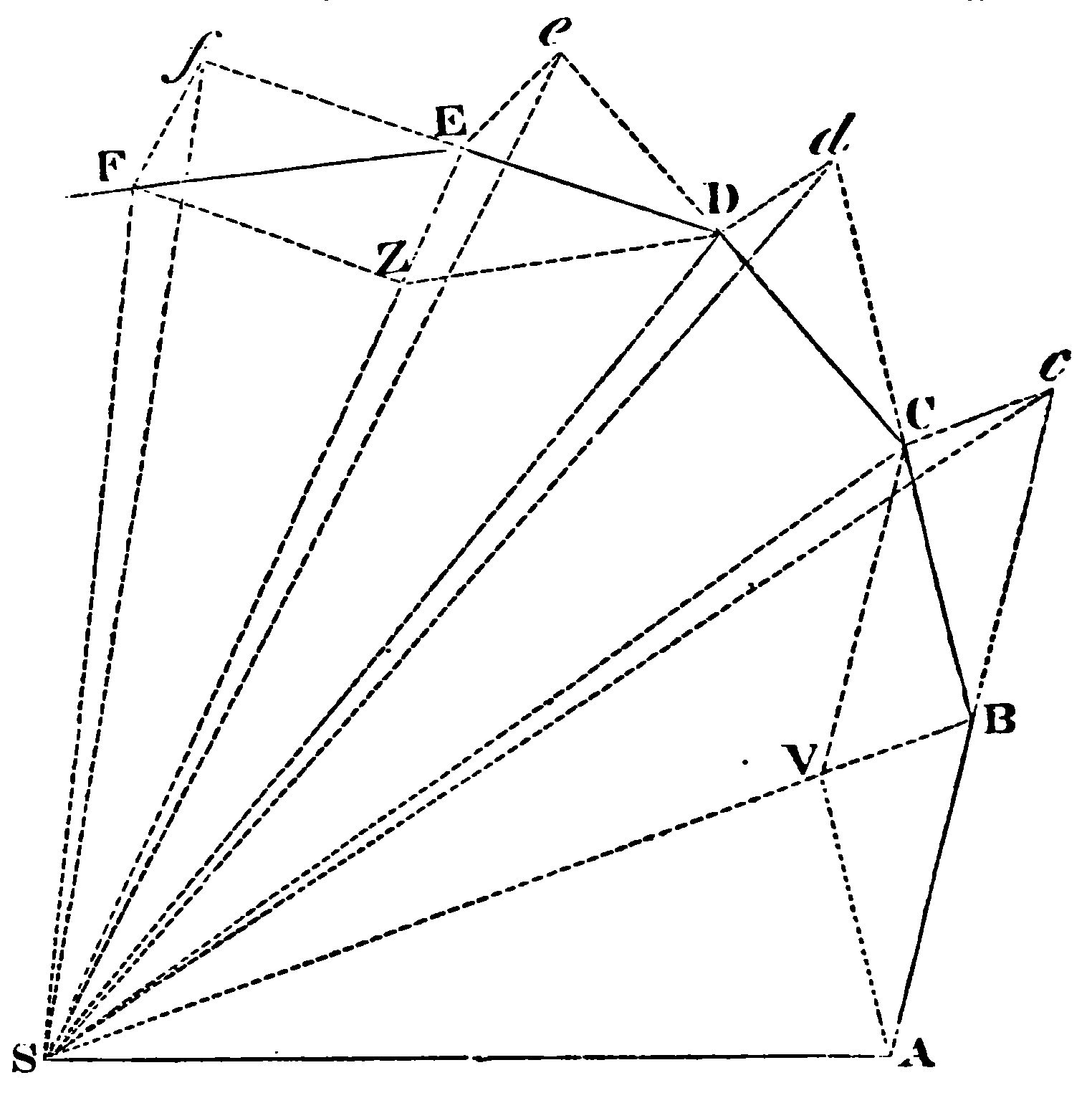}
    \includegraphics[width=0.50\linewidth]{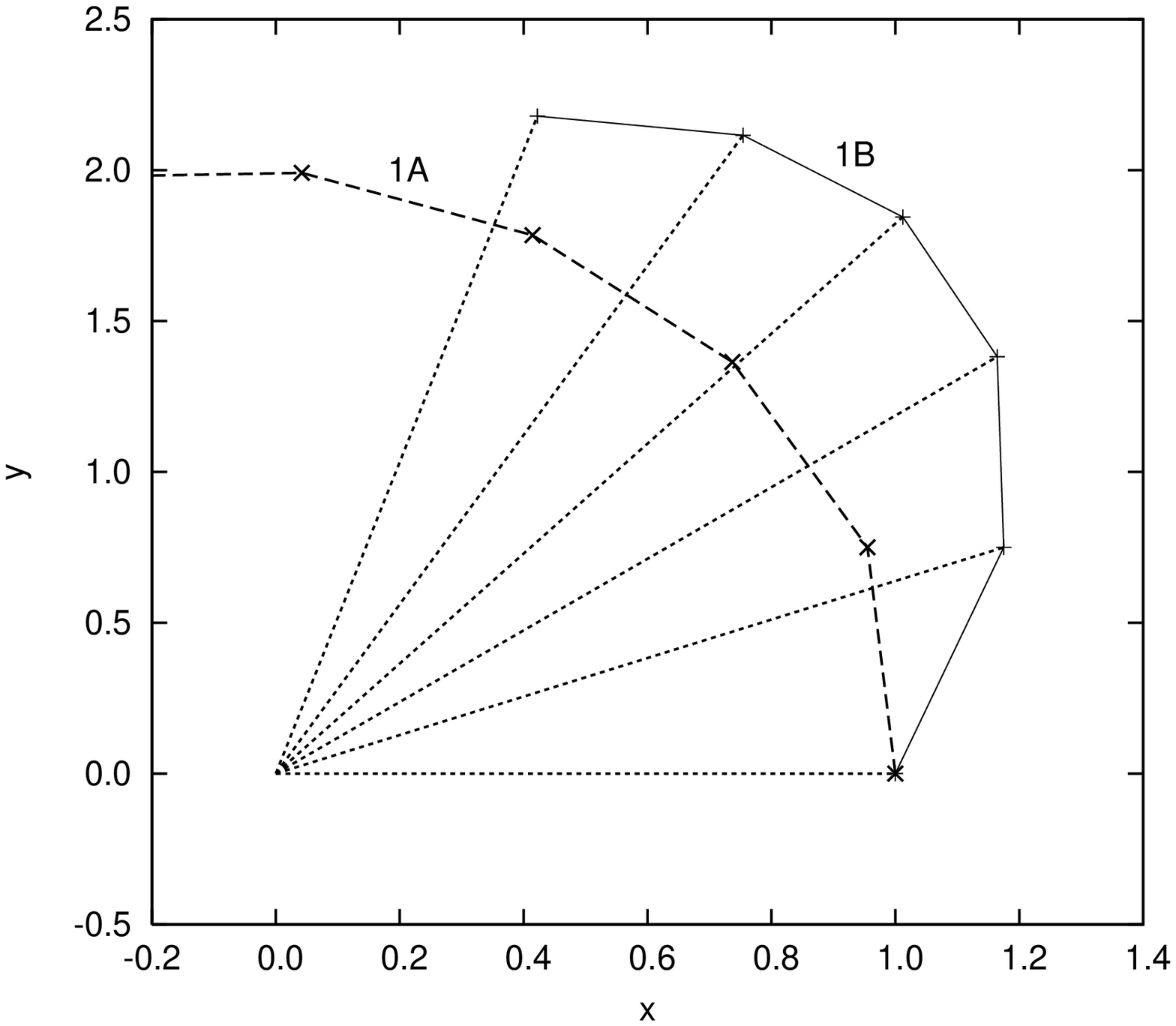}
	\caption{Left: Newton's Proposition 1 diagram taken from Ref.\onlinecite{prop1}.
Right: trajectories generated by algorithms 1A and 1B for a constant central force.
	} 
	\la{prop1}
\end{figure} 

We show also in Fig.1, the trajectory generated by algorithm 1A using the same
initial condition as 1B. It clearly does not match Newton's diagram. Algorithm 1A
corresponds to applying Newton's first two laws in the reversed order, producing
a trajectory starting with velocity $\bv_1$.

\section {Symplectic integrators and time reversibility}
\la{sym}

There have been continued debates over whether Newton's graphical method, 
which uses a series of {\it impulses}  $\dt\ac(\br_n)$, can converges to the 
continuum limit\cite{pou03}. 
All such discussions are mute because not only is Newton's graphical method a canonical transformation, 
it can also be rigorously derived  as a symplectic integrator from the  
Baker-Campbell-Hausdorff (BCH) formula 
\ba
\e^{\dt T} \e^{\dt V}=\e^{\dt(T+V)+\frac12 \dt^2[T,V]+\cdots},
\la{bch}
\ea
where $T$ and $V$ are operators:
\be
T=\bv\cdot \frac{\partial }{\partial \br}\qquad V=\ac(\br)\cdot \frac{\partial }{\partial \bv}. 
\la{cform}
\ee
More details can be found in Refs.\onlinecite{yos93,don05,scu05}, but for completeness, we can
give a brief summary here. 
Fundamental to Hamiltonian mechanics\cite{jor04} is Hamilton's equation
\begin{equation}
\dot q_i={{\partial H}\over{\partial p_i}},\qquad    
\dot p_i=-{{\partial H}\over{\partial q_i}}.
\la{hameq}
\end{equation}
and the idea that $q_i$ and $p_i$ are equally fundamental and {\it independent} dynamical variables:
$dp_i/dq_j=0$ and $dq_i/dp_j=0$. Consequently, if $v_i\equiv p_i/m$, then
\ba
\e^{\dt v_x\frac{\partial }{\partial x}}f(x,v_x)&=&(1+\dt v_x\frac{\partial }{\partial x}+\frac1{2!} (\dt v_x)^2\frac{\partial^2 }{\partial x^2}+\cdots)f(x,v_x)\nn\\
&=&(f+\dt v_x\frac{\partial f }{\partial x}+\frac1{2!} (\dt v_x)^2\frac{\partial^2 f }{\partial x^2}+\cdots)\nn\\
&=& f(x+\dt v_x,v_x).
\la{shift}
\ea
Therefore for any dynamical variable $f(\br,\bv)$, 
\ba
\frac{d}{dt}f(\br,\bv)=(\dot\br\cdot \frac{\partial }{\partial \br}+\dot\bv\cdot \frac{\partial }{\partial \bv})f(\br,\bv).
\nn
\ea
Invoking Hamilton's equation (\ref{hameq}) for the standard Hamilonian (\ref{ham}) gives,
\ba
\frac{d}{dt}f(\br,\bv)=(\bv\cdot \frac{\partial }{\partial \br}
+\ac(\br)\cdot \frac{\partial }{\partial \bv})f(\br,\bv)=(T+V)f(\br,\bv),
\nn
\ea
with solution
\be
f(\br,\bv,\dt)=\e^{\dt(T+V)}f(\br,\bv,0).
\nn
\ee
For $\dt$ small, one can then use (\ref{bch}) to approximate $\e^{\dt(T+V)}$ by $\e^{\dt T} \e^{\dt V}$.
The effect of $\e^{\dt T} \e^{\dt V}$ on any function $f(\br,\bv)$ is then, generalizing (\ref{shift}):
$$
\e^{\dt T} \e^{\dt V}f(\br,\bv)=\e^{\dt T}f(\br,\bv+\dt\ac(\br))=f(\br+\dt\bv,\bv+\dt\ac(\br+\dt\bv)),
$$
which is precisely equivalent to updating $\br'=\br+\dt\bv$ then $\bv'=\bv+\dt\ac(\br')$,
corresponding to algorithm 1B. (Note that the operators act from
right to left, but the resulting algorithm is equivalent to operators acting {\it sequentially} from
left to right.) Similarly, the action of $\e^{\dt V} \e^{\dt T}$ reproduces algorithm 1A.
Any decomposition of $\e^{\dt(T+V)}$ into a product of $ \e^{a_i\dt T}$ and $\e^{b_i\dt V}$,
with suitable coefficients $a_i$ and $b_i$, produces a symplectic algorithm.\cite{yos93,don05,scu05}

In the limit of $\dt\rightarrow 0$ the convergence of Newton's algorithm is guarantees 
by BCH:
$$\e^{\dt T} \e^{\dt V}\rightarrow \e^{\dt(T+V)}+O(\dt^2),$$
where the exact trajectory is given by $\br(t+\dt)=\e^{\dt(T+V)}\br(t)$
and $\bv(t+\dt)=\e^{\dt(T+V)}\bv(t)$.
There is therefore no question about the convergence, which one
can easily verify numerically, the only question is whether this
convergence is efficient, which we will discuss in the next section. 

From this operator form of 1B, it is well-known that it is 
not time-reversible, since
$$\e^{\dt T} \e^{\dt V}\e^{-\dt T} \e^{-\dt V}\neq 1.$$
(Recall that operators act sequentially from left to right.)
However, as we have shown in Section \ref{cano}, 
 the ``one and half" 1B, corresponding to
$\e^{\dt T} \e^{\dt V}\e^{\dt T}$
is time-reversible,
$$
\e^{\dt T} \e^{\dt V}\e^{\dt T}\e^{-\dt T} \e^{-\dt V}\e^{-\dt T}=1.
$$
It is obvious then that every left-right symmetric operator form as above
will cancel pair by pair and yield a time-reversible algorithm. However, recall also that 
the very similar Verlet algorithm, by eliminating the intermediate velocity,
is only time-reversible for every other position. 

In a recent publication, Nauenberg\cite{nau18} first transcribed Newton's graphical construction
in the Verlet form and claimed time-reversibility for Newton's Proposition 1. 
(From the Verlet form, he then inferred algorithm 1A, in contrasted to his earlier
deduction\cite{nau94} of 1B. This work supports his earlier identification.) 
Nauenberg's time-reversibility discussion in Ref.\onlinecite{nau18} is the time reversibility of
the Verlet algorithm, the reversibility of every {\it other} position on the trajectory,
but not {\it every} position on the trajectory. 

As discussed in Section \ref{cano}, Newton's Proposition 1 corresponds to algorithm 1B.
If that were completely the case, then as shown above, Newton's Proposition 1 would not 
be time-reversible. However, it is clear that Newton's intention was to compute {\it positions}
only and would have stopped after the last computation of the position. This interpretation would
imply that his graphic method is  a ``$n+1/2$'' 1B, or equivalently, a ``$1/2+n$'' 1A algorithm,
corresponding to,
$$
(\e^{\dt T} \e^{\dt V})^n\e^{\dt T}=\e^{\dt T} (\e^{\dt V}\e^{\dt T})^n,
$$ 
in which only the position but not the velocity is updated at the final
step. In this case, the above sequence of operators is left-right symmetric and Newton's
method would be completely time-reversible for every position of the trajectory. 

The algorithm corresponding to 
$$\e^{\dt V} \e^{\dt T}\e^{\dt T} \e^{\dt V},$$
by applying algorithm 1B and 1A alternately, is also time-reversible and
second-order in $\dt$. This is one form of the St\"ormer's algorithm\cite{storm},
at time-step $2\dt$.

In short, Newton's graphical construction is actually ``$n+1/2$'' 1B,  or ``$1/2+n$'' 1A,
totally time-reversible, not strictly 1B nor 1A, not equivalent to Verlet's nor 
St\"ormer's algorithm.

\section {The Error Hamiltonian}
\la{error}

By identifying Newton's proposition 1 as basically the symplectic algorithm 1B, one then knows
everything about this algorithm, and hence Newton's graphical construction. 
For example, the trajectory generated by the algorithm is {\it exact}
for the approximate Hamiltonian \cite{yos93,don05,scu05}
$$
H_A=H+\dt H_1+\dt^2H_2+\cdots
$$
where $H$ is the original Hamiltonian one is seeking to solve.
In the limit of $\dt\rightarrow 0$, only the first-order error Hamiltonian $H_1$ matters.
For algorithm 1B, this is well known to be \cite{yos93,don05,scu05}
$$
H_1=-\frac12\bv\cdot\ac(\br).
$$
Therefore in solving the linear central force problem with $\ac(\br)=-\br$, 
the algorithm is governed by the approximate Hamiltonian
\be
H_A=\frac12 (\bv^2+\br^2+\dt\bv\cdot\br),
\la{hare}
\ee
with a first order error Hamiltonian {\it less} singular than
the original Hamiltonian and only results in minor distortions.
It turns out, this first-order approximate Hamiltonian is
exactly conserved by algorithm 1B. (This was noted by Larsen\cite{lar83} 
for algorithm 1A, with a minus $\dt$ term, in 1983.)
Therefore the resulting trajectory of Newton's algorithm, 
as long as $|\dt|<2$, {\it is  always a closed ellipse}, even if it is not the correct 
elliptical orbit of the original Hamiltonian! See the Appendix for details. This is the reason
why Hooke, when using Newton's graphical method,
can claim that a linear force produces an elliptical orbit \cite{nau94}.

By contrast, for the inverse-square central force, the approximate
Hamiltonian is
$$
H_A=\frac12v^2-\frac1{r}+\frac{\dt}2 \frac{\bv\cdot\br}{r^3}.
$$
This first-order error Hamiltonian is {\it more} singular than the original Hamiltonian
(near the force center) and will cause the orbit to precess\cite{chin07} (and therefore not close),
with limited stability, unless $\dt$ is extremely small. Thus Hooke, when applying Newton's graphical
method to an inverse-square force, with a fairly large $\dt$, would not have been able to
produce a closed ellipse of moderate eccentricity.\cite{nau94} For the same reason, 
despite being the cornerstone of the {\it Principia}, Proposition 1
was never used, even by Newton, to prove an elliptical orbit from an inverse-square force.

\section {Conclusions}
\la{con}

In this work, we have shown that knowledge of canonical transformations and symplectic integrators,
not only help us to appreciate the prophetic nature of Newton's proposition 1, which anticipated symplectic
integrators centuries earlier, but also help us to understand the historical success and failure of Hooke's 
effort to determine central force orbits using Newton's graphical method.

\vskip .2in
\leftline{\bf APPENDIX}

\begin{figure}
	%	\vspace{0.5truein}
    \includegraphics[width=0.90\linewidth]{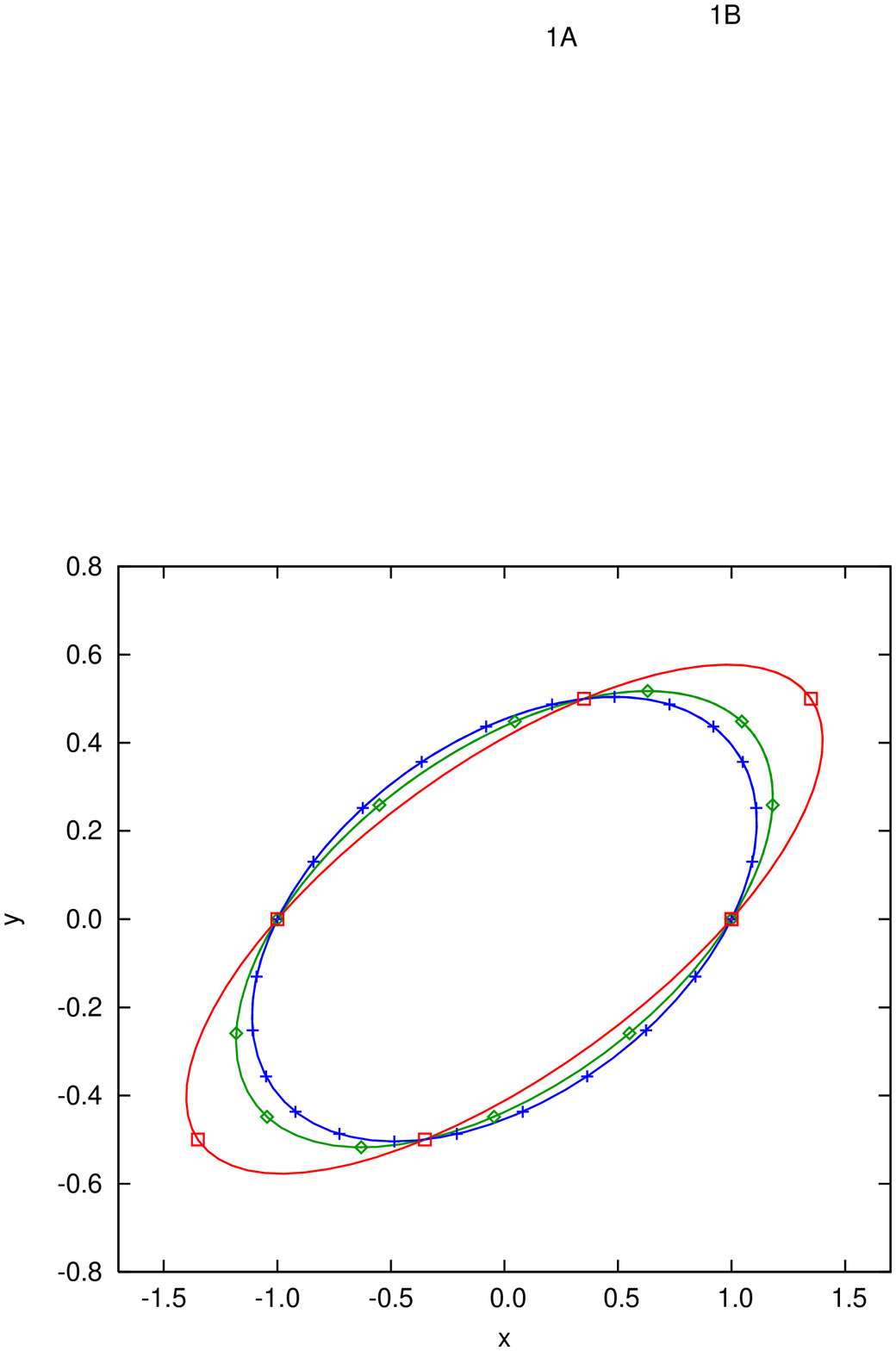}
	\caption{For a linear central force, Newton's proposition 1 always produces an ellipse, regardless of the time step size used. The symbols are the outputs of Newton's algorithm, the solid lines are
the predicted orbits (\ref{htraj}) of the algorithm. The three time-step sizes used are those of (\ref{step}).
	} 
	\la{ellip}
\end{figure}

The approximate Hamiltonian (\ref{hare}) can be rewritten as
$$
H_A=\frac12 \left(\bv'^2+(1-\frac{\dt^2}{4})\br^2\right),
$$
with $\bv'=\bv+\dt\br/2$. Therefore as long $|\dt|<2$, it is a harmonic oscillator with
angular frequence $\omega=\sqrt{1-\dt^2/4}$ and trajectory
\be
\br(t)=\br_0\cos(\omega t)+\frac{\bv'_0}{\omega}\sin(\omega t)
\la{htraj}
\ee
where $\bv'_0=\bv_0+\dt\br_0/2$. While (\ref{htraj}) gives the correct trajectory
of the algorithm, the algorithm's approximate angular frequence is {\it not} $\omega$ given
above, but is given by\cite{scu05b}
$$
\omega_A=\frac1{\dt}\cos^{-1}(1-\dt^2/2).
$$
One can therefore choose $\dt$ so that $T/\dt=(2\pi/\omega_A)/\dt$ is an integer.
For $T/\dt=6, 12, 24$, one requires
\be
\dt=1,\sqrt{2-\sqrt{3}}\approx 0.51764, \sqrt{2-\sqrt{2+\sqrt{3}}}\approx 0.26105, 
\la{step}
\ee
respectively.
In Fig.\ref{ellip}, we compare the output of Newton's algorithm at these three time steps
with the analytical orbit of (\ref{htraj}). For the linear central force case,
Newton's algorithm always produces positions which are exactly on an ellipse, 
as long as $|\dt|<2$. (At $|\dt|=2$ the orbit collapses into a line.) At a large $\dt$,
as in the case of $\dt=1$, these ellipses are far from the correct orbit, as
illustrated in Fig.\ref{ellip}.

\begin{acknowledgments}
I thank my colleague Wayne Saslow, for calling my attention to Ref.\onlinecite{nau18}, which
inspired this work.
\end{acknowledgments}

%\newpage

%%%%%%%%%%%%%%%%%%%%%%%%%%%%%%%%%%%%%%%%%%%%%%%%%%%%%%%
\end{document}